\documentclass{article}

\PassOptionsToPackage{numbers, compress,sort}{natbib}
\usepackage[preprint]{neurips_2020}

\usepackage[utf8]{inputenc} 
\usepackage[T1]{fontenc}    
\usepackage{hyperref}       
\usepackage{url}            
\usepackage{booktabs}       
\usepackage{amsfonts}       
\usepackage{nicefrac}       
\usepackage{microtype}      
\usepackage[inline]{enumitem}
\usepackage{amssymb}
\usepackage{natbib}
\usepackage{graphicx}
\usepackage{url}
\usepackage{epigraph}
\usepackage{lscape}
\usepackage{multirow}
\usepackage{color, colortbl}

\newcommand{\negskip}{\vspace*{-.5\baselineskip}}

\definecolor{Gray}{gray}{0.9}

\author{%
  Svitlana Vakulenko\\
  University of Amsterdam\\
  Amsterdam, The Netherlands \\
  \texttt{s.vakulenko@uva.nl} \\
  \And
  Vadim Savenkov \\
  Vienna University of Economics and Business \\
  Vienna, Austria \\
  \texttt{vadim.savenkov@wu.ac.at} \\
  \AND
  Maarten de Rijke \\
  University of Amsterdam \& Ahold Delhaize \\
  Amsterdam, The Netherlands \\
  \texttt{m.derijke@uva.nl} \\
}

\title{Conversational Browsing: A Preliminary Study}
\title{Conversational Browsing: An Exploration}
\title{Conversational Browsing}

\begin{document}

\maketitle

\begin{abstract}
    How can we better understand the mechanisms behind multi-turn information seeking dialogues? 
    How can we use these insights to design a dialogue system that does not require explicit query formulation upfront as in question answering?
    To answer these questions, we collected observations of human participants performing a similar task to obtain inspiration for the system design.
    Then, we studied the structure of conversations that occurred in these settings and used the resulting insights to develop a grounded theory, design and evaluate a first system prototype.
    Evaluation results show that our approach is effective and can complement query-based information retrieval approaches.
    We contribute new insights about information-seeking behavior by analyzing and providing automated support for a type of information-seeking strategy that is effective when the clarity of the information need and familiarity with the collection content are low.
\end{abstract}

\section{Introduction}
\label{section:intro}


Conversational search has evolved as a new paradigm with the goal of making information retrieval interfaces feel more natural and convenient for their users~\citep{DBLP:conf/chiir/RadlinskiC17}.
Ongoing research and development efforts in this direction are now heavily skewed towards the question answering task~\citep{DBLP:conf/chi/VtyurinaSAC17,DBLP:conf/acl/RajpurkarJL18,DBLP:conf/emnlp/SaeidiBL0RSB018}.
However, there is ample evidence that conversational search interfaces need to support a more diverse set of interactions to be able to assist their users~\citep{DBLP:journals/corr/abs-1812-10720,bates1989design,belkin2016people}.
The limitation of the question answering interaction paradigm is in its inherent bias towards knowledge that the user already has: users need to be able to formulate an appropriate question before they can engage with a question answering interface in a meaningful way~\citep{belkin1982ask}.
A similar issue occurs also in situations when a system poses questions to its user~\citep{DBLP:conf/cikm/ZhangCA0C18}.

In this paper we focus on the task of information presentation in conversational search interfaces designed to communicate all available knowledge to the user.
While similar to the information presentation task required for traditional spoken dialogue systems, which list available options in response to a user query~\citep{DBLP:conf/eacl/DembergM06}, what we have in mind goes beyond this paradigm.
\citet{DBLP:journals/corr/abs-1709-05298} have proposed to apply the concept of \emph{interactive storytelling}~\citep{DBLP:conf/icids/2018} to conversational search systems for exploratory search~\citep{DBLP:series/synthesis/2009White}.
Interactive storytelling is an extension of computational storytelling that makes the story generation process dynamicaly adopt in response to the user input~\citep{DBLP:journals/aim/RiedlB13}.
We are interested in applying interactive storytelling to explore the content of an information source thereby forming a conversation between a user and a system.
Conversational exploratory search can be useful in a range of knowledge access scenarios, including education and e-commerce.
For example, an intelligent shopping assistant should be able to fluently guide a customer through the whole product catalog, carefully picking up on the user's reactions to form a preference model and adopt the exploration direction that optimizes customer satisfaction.

We view \emph{conversational browsing} as a first step towards the bigger agenda of enabling conversational exploratory search via interactive storytelling~\citep{DBLP:journals/corr/abs-1709-05298}.
It focuses primarily on supporting navigation control, where a user can influence and change the direction of exploration.
Conversational browsing is a task designed to enable conversational exploratory search for structured information sources, such as a database table or a knowledge graph.
The goal of this conversational interface is to unfold the content of the collection to the user in an interactive manner, that is, in response to their chosen exploration direction.
Explicit structure of an information source allows us to model it as a graph abstraction and evaluate different navigation strategies, i.e., the sequence in which the nodes can be visited.

We start with the basic setup in which an ``interactive story'' is to be generated from a single database table.
Our main research question is how to enable efficient information access in a situation where the information goal of the user is implicit or vaguely defined.
Examples include cases in which the user is not familiar with the domain vocabulary, wants to understand the available content and structure of the information source, or is simply looking for inspiration and serendipitous discovery.
In this paper we describe and evaluate the design of an automated dialogue system that helps users to acquire knowledge about the structure and content of a catalogue through dialog-based interaction without the requirement to specify their information need in advance.
The kind of system we have in mind has to be considerate of the user, in the sense that it should account for:
\begin{itemize}
\item \textit{cognitive load} to determine and regulate a reasonable pace of the information flow; and
\item \textit{user preferences} for the user to regulate the direction of the information flow, i.e., conversation topic.
\end{itemize}

To be able to formulate an informed hypothesis of what kind of interaction the envisaged system should provide we seek to get inspiration from human-to-human conversations collected in a controlled laboratory study.
To this end, we follow an end-to-end methodology from collecting and analyzing dialogue transcripts through model design, implementation and evaluation.
Our empirical study of the information-seeking dialogues and strategies that humans employ to communicate content of an information source informs the design of a \emph{conversational browsing model}.
This model describes a general information-seeking process and is applicable across different use cases, in contrast to  supervised models trained directly on dialogue transcripts~\citep{DBLP:journals/taslp/RieserLK14}.
We design a prototype based on the main concepts of our conversational browsing model and evaluate the prototype in a user study by contrasting it against a traditional conversational search system that follows request-response paradigm.
Thus, our conversational browsing model provides a general framework that not only provides a theoretical understanding of an information-seeking dialogue but also forms the basis for system design.

In summary, our main contributions in this paper are:
\begin{itemize}
\item a \textbf{dataset} of dialogue transcripts that provide insights on human strategies in information-seeking conversations, in which an information provider takes on a pro-active role;
\item a \textbf{model} that systematizes these insights as a set of requirements, components and functionality they should support to automate such information-seeking conversations; and
\item an \textbf{evaluation} of a proof-of-concept implementation of a conversational browsing system.
\end{itemize}

We find that conversational browsing can be a powerful tool able to mitigate the vocabulary mismatch problem and assist search.
Based on the conversational data that we collect, we discover that the essence of conversational browsing interaction lies in the recurrent process of vocabulary exchange that attempts to iteratively reduce the search space and bring an information seeker closer to their information goal.

We proceed by briefly reviewing the major theories of information-seeking strategies and introduce the concepts of exploratory search and interactive storytelling in Section~\ref{section:background}.
Section~\ref{section:methodology} provides an overview of the methodology that we followed to design and evaluate our conversational browsing system.
For our empirical evaluation and data collection steps we instantiated the task of conversational browsing with a concrete use case described in Section~\ref{section:usecase}.
We proceed by describing the setup and the outcomes of a user study organized to collect a dataset of human-to-human conversations (Section~\ref{section:dataset}), which served as a blue-print for our dialogue system design (Section~\ref{section:design}). 
Finally, we describe the evaluation of our conversational browsing prototype in Section~\ref{section:evaluation}.
We consider two evaluation setups: (1)~a user simulation, which helps us to tune the hyper-parameters of the model and estimate the system performance, and (2)~a user study to test our modeling assumptions with target users.
We conclude by discussing the relation of our findings to previous work in Section~\ref{section:related}.

\section{Background}
\label{section:background}

% ...

\subsection{Information seeking}

Information-seeking behavior is a complex process that has been extensively studied in the literature and several models have been proposed as an attempt to describe its structure and   characteristics~\citep{bates1979information,ellis1989behavioural,kuhlthau1991inside}.
A search session is an instance of this process, which may include several interaction turns (information exchanges) between the user and a search engine as an information source~\citep{hagen2011query}.
The \emph{berry picking} model of search behavior suggests that the understanding of an information need evolves during the search process as more information becomes available to the seeker and browsing interfaces are the key enablers for supporting this kind of a dynamic search process~\citep{bates1989design}.

One of the major challenges in information seeking is query formulation~\citep{kelly2009comparison}.
The concept of the anomalous state of knowledge (ASK) suggests that it is not reasonable to expect ``that it is possible for the user to specify precisely the information that she/he requires.''~\citep{belkin1982ask}
Recognition of this phenomena is an important step towards considering alternative solutions that may help information seekers in practice.

``Naturally occurring'' conversations have often served as evidence and the source of inspiration for developing theoretical models and prototypical implementations~\citep[see, e.g.,][]{trippas2016how,DBLP:journals/corr/abs-1812-10720,DBLP:conf/interspeech/SegundoMCGRP01}.
For example, observations of information-seeking dialogues with a reference librarian, which is a classic example of a help-desk information service, suggest that a process of negotiation is an important component that helps to better understand and adjust the information need when discussing it with an expert~\citep{taylor1968}.
The key properties of an information-seeking dialogue are asymmetry of roles and cooperation --- the information seeker and provider (intermediary) cooperate to better understand and satisfy the information need of the seeker, i.e., the information provider has access to the information source and plays the role of an intermediary for the information seeker who has an information need.
Here is an excerpt from a real chat-based conversation given below, which occurred between a pair of students during a lab study that we conducted, which is described in more detail in Section~\ref{section:dataset}.
One of the students seeks information (S -- Seeker) and the other one is trying to help using the Austrian Open Data portal (I -- Intermediary):
\begin{enumerate}[nosep]
\item[\textbf{(I)}] \url{data.gv.at} is an Austrian Open Data portal.
\item[\textbf{(S)}] What kind of data can you find there?
\item[\textbf{(I)}] You can search for datasets in economics and politics categories, but also education, sports, culture etc.
\item[\textbf{(S)}] What exactly do you mean?
\item[\textbf{(I)}] Statistics about birth rates, kindergartens locations, public transport, for example.
\item[\textbf{(S)}] What data do you have related to birth rates?
\item[\textbf{(I)}] I can tell you statistics about the places where newborn live or the names they get.
\item[\textbf{(S)}] That sounds all right. I'm curious about the names...
\end{enumerate}
We collected and analyzed this kind of dialogue transcripts (see Section~\ref{section:dataset}) to come up with a general framework, which we formalized into a conversational browsing model.
Our experiments show that a dialogue system design based on this model is effective in providing basic exploratory search (browsing) functionality.

\subsection{Conversational exploratory search}

The goal in exploratory search is to provide guidance for seekers who are exploring unfamiliar information landscapes~\citep{Marchionini:2006:Exploratory,DBLP:series/synthesis/2009White}.
\citet{DBLP:series/synthesis/2009White} distinguish between two main activities within the exploratory search paradigm: \emph{exploratory browsing} and \emph{focused searching}.
Exploratory browsing is an initial step that provides necessary domain understanding required for focused searching activities.
It is related to \citet{DBLP:conf/chiir/RadlinskiC17}'s \emph{system revealment} property: ``The system reveals to the user its capabilities and corpus, building the user's expectations of what it can and cannot do.''
Browsing is one of the information-seeking activities that is defined as ``semi-directed or semi-structured searching"~\citep{ellis1989behavioural}, i.e., the information need is vague and the goals include general collection understanding and serendipitous discovery~\citep{mckay2017manoeuvres}.
The purpose of browsing interface design is to make aspects of the collection apparent to the user and provide the means to traverse (navigate) between different options, which requires making the choice of the interaction modes and definition of a closed set of eligible operations available to the system and its user in advance~\citep{DBLP:conf/semweb/NunesS15}.

\begin{figure*}
\includegraphics[clip,trim=0mm 15mm 0mm 0mm,width=\textwidth]{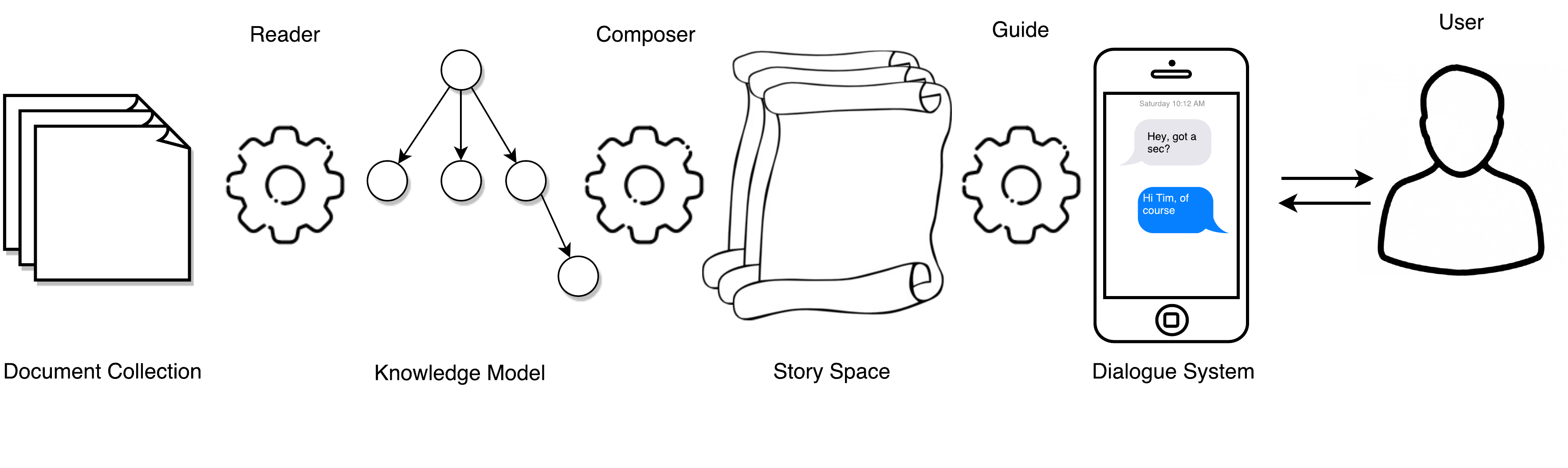}
\caption{Communicating knowledge via an interactive storytelling process.}
\label{fig:architecture}
\end{figure*}

Conversational agents and search systems are becoming increasingly popular~\citep{DBLP:conf/chi/VtyurinaSAC17}.
However, such systems mainly focus on question answering and simple search tasks, i.e., those that are to a large extent solved by web search engines.
Conversational agents and search systems should also support exploratory search~\citep{DBLP:journals/corr/abs-1709-05298}.
A conversational exploratory search system is represented in Figure~\ref{fig:architecture}.
It has a number of key components: Document Collection, Knowledge Model, Story Space, Dialog System and User.
These components are connected through the Reader, Composer, and Guide modules.
The interplay of the system components and modules happens at different stages.

\subsubsection{Knowledge representation} 
Knowledge representation consists of a \textit{Reader} module that extracts concepts and relations from the Document Collection and embeds them into a single Knowledge Model.
The Knowledge Model integrates different elements (words, concepts or entities) and describes relations between them.
The knowledge can be explicitly modeled by means of a taxonomy or ontology (knowledge graph) but it can also be embedded into a latent structure.

\subsubsection{Story generation} 
Story generation consists of the \textit{Composer} module that is able to generate stories by combining elements of the Knowledge Model.
To create a story, the Composer has to select elements (characters, words, facts, concepts, relations), choose their ordering, arrange selected elements in time and/or space.
The set of all possible stories constitutes the Story Space.

\subsubsection{Interactive storytelling} 
Interactive storytelling includes a \textit{Guide} module that helps the User to navigate through the Document Collection via the Story Space.
The Guide can change the current position within a single story or traverse the space across different stories.
Interactive storytelling integrates the Dialogue System to communicate a story to the User and to receive an input from the User. Supporting such a conversation with the User requires natural language (utterance) generation and understanding. Note that the input/output modalities do not have to be restricted to text and speech only and may include images, videos, interactive visualization, virtual reality interactions, etc.






\medskip\noindent%
A conversational exploratory search system should support the following types of the user-system interactions:

\begin{itemize}

\item \textit{Navigation Control} -- a user chooses a direction (branch) for exploration and is also able to influence and change the current direction of the narrative at any point in time;

\item \textit{Feedback} -- a user may provide feedback to the system (positive, neutral, negative) that may help to correct and steer the direction of the story that shall maximize the user satisfaction with the system;

\item \textit{Question} -- a user may pose questions to the system, e.g., a request for a definition, look up query, etc.

\end{itemize}

In this paper we focus on mechanisms for navigation control in conversational browsing, which is a version of conversational exploratory search specifically designed for structured information sources, such as databases and knowledge graphs.
We ground our model design in observation of information-seeking dialogues paying special attention to the ability of human information providers to prioritize and structure information into coherent chance to balance the cognitive load of an information seeker.

\subsection{Cognitive load}
\textit{Cognitive information processing} (CIP) theory~\citep{atkinson1968human} is a popular model describing diverse cognitive processes.
The central idea in CIP is that the human mind can be modeled as an information processor that receives information from the senses (input), processes it, and then produces a response (output).
Learners are viewed as active seekers and processors of information.
CIP focuses mainly on the information processing task, in particular the aspects of memory encoding and retrieval.

\textit{Cognitive load theory} (CLT)~\citep{sweller1988cognitive} builds upon the information processing model.
``Cognitive load'' relates to the amount of information that working memory can hold and operate on at one unit of time.
The capacity of the human working memory is very limited.
When too much information is presented at once, it becomes overwhelmed and much of that information is subsequently lost.
CLT aims at making learning more efficient, calling for communication strategies that take into account cognitive limitations of the human mind.

\textit{Span theory}~\citep{bachelder1977theory} is a behavioral theory describing the relation between performance and span load, a fundamental task characteristic. In particular, several researchers studying the limits of human cognitive abilities point to the average number of objects a human brain can hold in working memory, i.e. the working memory capacity. 
The famous ``magic number'' originally suggested by \citet{miller1956magical} was $7\pm 2$ objects, while more recent research shows that this estimate is too optimistic and suggests the new limit close to 4~\citep{cowan_2001} objects.

One other process that seems to be limited at about 4 elements is subitizing, the rapid enumeration of small numbers of objects. When a number of objects are flashed briefly, their number can be determined very quickly, at a glance, when the number does not exceed the subitizing limit, which is about 4 objects.
Larger numbers of objects must be counted, which is a slower process. 
 
However widely criticized as a single number not reflecting the task difficulty and individual differences, this number is supported by a remarkable degree of similarity in the capacity limit of working memory observed in a wide range of procedures and is likely to reflect a reasonable distribution mean able to inform chunking decisions for efficient information processing by humans. 
Research also shows that the size, rather than the number, of chunks that are stored in short-term memory is what allows for enhanced memory in individuals.
A chunk is defined as a collection of concepts that have strong associations to one another and much weaker associations to other chunks acting as a coherent, integrated group~\citep{cowan_2001}.
It is believed that individuals create higher order cognitive representations of the items on the list that are more easily remembered as a group than as individual items themselves.

Over the years various readability formulas have been proposed to predict comprehension difficulty of a text passage~\citep{crossley2017predicting}.
For example, one of the early studies~\citep{mc1969smog} reports a negative correlation between reading efficiency and the count of polysyllable words, defined as words of three or more syllables.

The combination of linguistic features capturing two elements of text difficulty (lexical and syntactic complexity) constitutes a good predictor for the time required to process text and comprehension~\citep{crossley2017predicting}, e.g.,
\begin{itemize}
\item texts are more comprehensive if the words are less sophisticated, there are fewer verbs, and lower text cohesion; and
\item a larger number of unique trigrams and proper nouns per sentence slows processing
\end{itemize}
Another important set of features that describe the intrinsic properties of the reader's mental model, as an essential part and a prerequisite for the success of the text comprehension process, include individual reading ability, background knowledge, etc.

We draw upon the results in cognitive science to inform our model and design an effective dialog system for conversational exploratory search.
More specifically, we explore the structure of messages exchanged in an information-seeking conversation -- the number and relations between the concepts contained within a single message, as part of an important mechanism designed to support human information processing abilities with respect to the cognitive properties of a human mind.

\section{End-to-end Methodology for Designing a Dialogue System}
\label{section:methodology}

\begin{figure}[t]
\centering
\includegraphics[width=.6\textwidth]{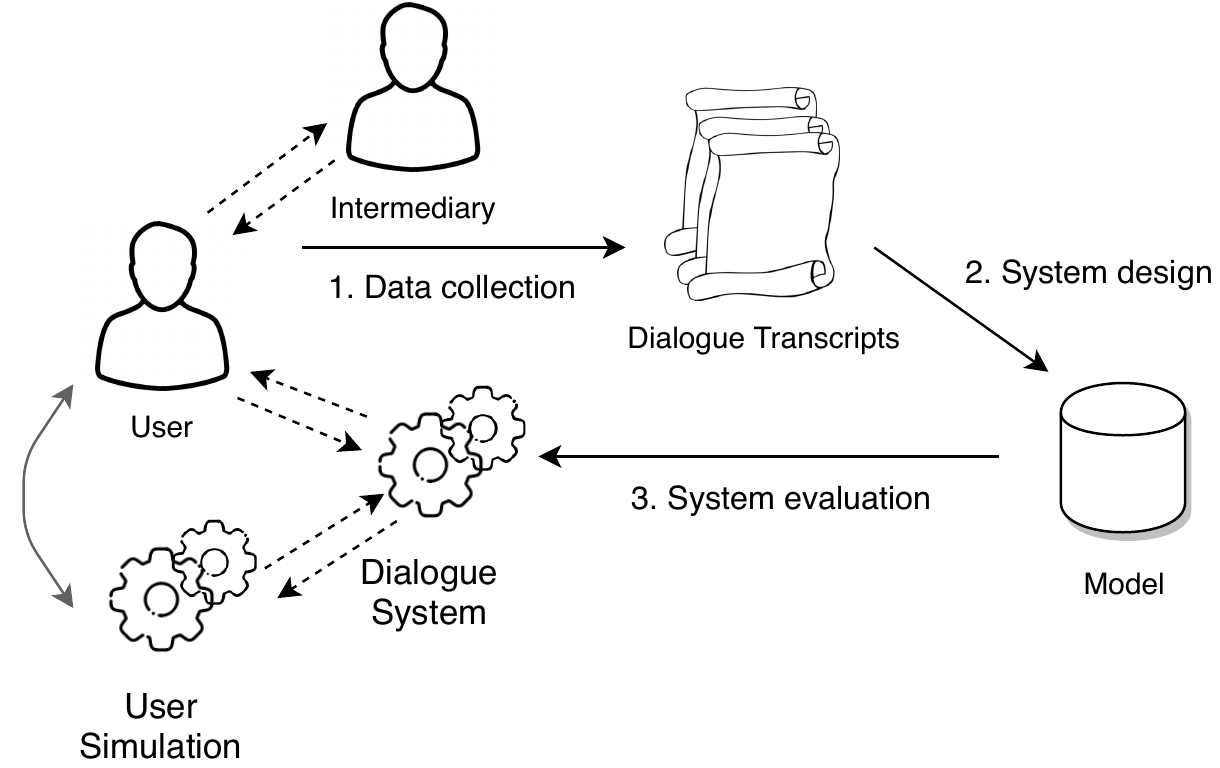}
\caption{End-to-end methodology for designing a dialogue system.}
\label{fig:methodology}
\end{figure}

We follow an end-to-end methodology for designing a dialogue system outlined in Fig.~\ref{fig:methodology}.
It is a data-driven approach that helps us to formulate a general theoretical framework for conversational browsing and demonstrate its effectiveness on a sample scenario of open data exploration. 
We follow a mixed method approach to structure the design and evaluation of our conversational browsing system.

We start with a user study to collect samples of conversations and learn from the strategies human participants employ in the context of a conversational browsing task, which results in a dataset of dialogue transcripts (1.~Data collection). 
These empirical observations inform the design of a theoretical framework for conversational browsing that we propose in this paper (2.~System design). 
The model is formally defined on a higher level of abstraction to separate the details inherent for the particular scenario from the general structure, which can be reused across other use cases.
In the final stage (3.~System evaluation), we implement a prototype of a conversational browsing interface following the proposed framework and evaluate it via: (1) a user simulation; (2) a laboratory study with real users~\citep{Koren:2008:PIF:1367497.1367562}.
We start by estimating an expected number of turns and tune the hyper-parameters in a user simulation.
Then, we conduct a second user study with real users to evaluate the assumptions behind our model's design.

\section{Use Case: Open Data \& Dataset Search}
\label{section:usecase}

This research was conducted in the context of CommuniData project\footnote{\url{https://www.communidata.at}} that focused on making open data, i.e. publicly available datasets, more accessible for lay users to support citizen participation and transparency in decision making.
Making data public does not equate enabling citizens to make use of the data~\citep{DBLP:conf/cedem/BenoFUP17}.

Dataset search evolved as a new research direction to address the challenges of data heterogeneity~\citep{DBLP:journals/sigir/KoestenMGSR18,DBLP:conf/chi/KoestenKTS17}.
Google Dataset Search aggregates metadata of 14M datasets from 3k repositories.\footnote{\url{https://toolbox.google.com/datasetsearch}}
One of the first features users asked for was to extend the interface beyond a search box to support a browsing functionality~\citep{DBLP:conf/www/BrickleyBN19}.
A recent study of open data portal logs~\citep{10.1007/978-3-319-60131-1_29} also highlights that open data sources ``are used exploratively, rather than to answer focused questions.''

Similarly, we also started earlier by designing a simple conversational search interface, which allows a user to submit a query and returns a list of matching datasets~\citep{DBLP:conf/esws/NeumaierSV17}.
The chatbot attracted a lot of attention but many users were not able to formulate adequate queries since they were not aware of the structure and terminology of the underlying repository.
In this case such conversational search interfaces turn out to be of little use.
This motivated us to look for alternative types of interaction that could allow users to understand which information a system can provide.




We formulated this task as \emph{conversational browsing}, which is reminiscent of web browsing that allows users to explore a vast web graph by traversing links between pages.
However, use cases for the conversational browsing functionality are not limited to the dataset search/exploration scenarios.
This task is relevant to other applications, in which a conversational interface relies on a large-scale structured information source, e.g., a database, table or index. 
Similar use cases arise in product search or content recommendation scenarios, when a user is willing to learn about the underlying structure of the collection and explore the alternatives to get a comprehensive overview of the available options.

\section{Data Collection}
\label{section:dataset}


The task was completed by 26 student participants grouped in pairs.
The volunteers were recruited among the undergaduate students taking a Data Processing course in our university.
All participants had previous experience with web search interfaces, but no previous experience with the web-sites they were instructed to access during the user study.
Every participant was assigned an individual working place in a lab equipped with a desktop PC. 
The pairs were seated apart from each other and had a separate web-based chat-room with a full-screen chat window, which served as the only channel for communication.\footnote{\url{https://tlk.io/}}
The conversational data\-set that we collected consists of 26 conversation transcripts with a total of 416 messages (minimum: 6, maximum: 36, mean: 16 messages per conversation).\footnote{\url{https://github.com/svakulenk0/ODExploration_data}}

\subsection{Task description}
\label{sec:task}

We designed two different tasks using two open data portals with faceted search interfaces so that every participant would have a chance to play both roles.
For one of the two portals they assumed the role of the Intermediary (I), and for the other the role of the Seeker (S). 
Seekers were assigned specific information goals (to find one of the datasets from the portal) but were explicitly instructed not to share the goal with the Intermediary but try to reach it by providing relevant feedback to the Intermediary, that is, choosing from the relevant exploration directions suggested by the Intermediary.
Open data portals are particularly suited for such an experiment, since they provide a ready-made user search interface as well as a machine-readable API to access the same data repository, both publicly available.

The experimental procedure consisted of two phases: 
\begin{enumerate}
\item a learning phase, and 
\item a teaching (or knowledge sharing) phase.
\end{enumerate}
After each phase a quiz was completed to assess the acquired knowledge.

During the learning phase the Intermediary studied the structure and content of the collection using the website of the open data portal, which provides a faceted search interface. 
The Intermediary completed the quiz designed to evaluate the extent of the acquired knowledge about the structure and content of the collection by browsing the web site.
After completion of the learning phase, the Intermediary shared the acquired knowledge with the Seeker in a conversation. 
The Seeker completed a quiz to assess the extent to which they acquired knowledge about the structure and content of the collection through dialogue interactions with the Intermediary.

\subsection{Dataset description}
\label{sec:conversation_structure}

The conversational data\-set that we collected in this manner consists of 26 conversation transcripts with a total of 416 messages (minimum: 6, maximum: 36, mean: 16 messages per conversation).
Most of the conversations (24 out of 26) were identified as successful based on the Seeker's explicit feedback and the correct dataset link provided by the Intermediary by the end of a conversation.
However, little additional knowledge beyond the specific information goal provided in the task was shared between the study participants (Seekers indicated that the topics were not discussed in the conversation).

One of the authors annotated each message in every transcript with the speaker identifier (a Seeker or an Intermediary role), clustered similar utterances and annotated them with labels reflecting the function they play in the conversation (dialogue act types), e.g., greeting or question.
In total, we identified 15 distinct utterance types in this dataset shared across different conversations (see the full list with descriptions in Table~\ref{dialog_acts}).
Messages were also annotated with span-level labels to keep the count for the number of concepts per message and their types.

To understand the structure of the conversations collected in the dataset and relations between different dialogue acts, we extracted a model of the conversation flow from the conversation transcripts by feeding the transcript as sequences of annotated utterances into the ProM framework\footnote{\url{http://www.promtools.org}}~\citep{DBLP:books/sp/Aalst16}, which is a popular process mining software toolkit. 
The directed graph of the conversation model was constructed and visualized using the Inductive Visual Miner ProM plugin~\citep{DBLP:conf/bpm/LeemansFA14}. 
We provide a snapshot of the core of the extracted conversation process model (see \figurename~\ref{fig:conversation_model}), which describes the information exchange loop used by the conversation partners to traverse the information model in the direction of the information goal.

Many conversations begin with a ``hand-shaking'' message exchange, which may include mutual greetings, introductions and goal statements that provide the context for the rest of the conversation. 
For example, for the Seeker it would include a general question about the content of the information source, while for the Intermediary it would be a short description of the information source, an offer of the information service and a request about the concrete information goal of the Seeker. 

Conversations are mostly a vocabulary exchange aimed at traversing the information space towards the subset of items containing the information goal. 
The Intermediary (\textbf{I}) lists keywords, which correspond to a set of related concepts in the information space, and the Seeker (\textbf{S}) chooses one or more of these concepts to continue exploration. 
To illustrate, here is a snippet of a sample conversation:

\begin{enumerate}[nosep]
\item[\textbf{(I)}] It is an Open Data source that contains data about various \textbf{topics}: \textit{work}, \textit{culture}, \textit{education}, \textit{population}, \ldots\ What would interest you the most?
\item[\textbf{(S)}] \textit{Population}.
\item[\textbf{(I)}] Okay, is there a specific \textbf{region} for which you would like to find a dataset (\textit{Steiermark}, \textit{Vorarlberg}, \textit{Vienna} etc.)?
\item[\textbf{(S)}] I'm interested in the \textit{population} of \textit{Upper-Austria}.
\end{enumerate}

\begin{landscape}
\begin{table*}[tb]
\caption{Types of utterances exchanged between the agent A (Intermediary) and user U (Seeker).}
\label{dialog_acts}
\mbox{}\vspace*{-\baselineskip}
\centering
\begin{tabular}{ l rr l l l }
\toprule
\bf Type       & \multicolumn{2}{c}{\bf Occurrences}  & \bf Description                                                        & \bf Example                                                          & \bf Role \\
\midrule

list(keywords)   & 102 & (24.5\%) & Suggest available options for an exploration              & We have data on culture, finance. & A    \\
&&& direction & \\
set(keywords)    & 85 & (20.4\%)  & Choose an exploration direction                                    & I am interested in culture.                     & U    \\
confirm()        & 51 & (12.3\%)  & Confirm an exploration direction                                   & That would be perfect!                                            & UA \\
success()        & 36 & (8.7\%)   & Indicate reaching an information goal                             & Thank you very much!                                             & UA \\
question(data)   & 31 & (7.5\%)   & Indicate a general information need                               & What data do you have?                                           & U    \\
prompt(keywords) & 26 & (6.3\%)   & Request to specify the information goal                            & Any state that interests you?                   & A    \\
reject()         & 19 & (4.6\%)   & Reject an exploration direction                                    & This is not what I am looking for.               & U    \\
greeting()       & 19 & (4.6\%)   & Common start of the conversation                                   & Hello!                                                            & UA \\
bool(data)       & 13 & (3.1\%)   & Report whether requested subset exists & Yes, we have data about this year.~~~\mbox{}                                & A    \\
count(data)      & 13 & (3.1\%)   & Report the size of a subset                                        & There are 314 datasets in CSV.                            & A    \\
link(dataset)    & 9 & (2.2\%)    & Report link to a dataset                                           & There you go: \url{http://data}                      & A    \\
verify()         & 4 & (1.0\%)    & Prompt to confirm an exploration direction                        & Is that what you are looking for?                                & UA \\
more()           & 3 & (0.7\%)    & Request to continue in the same exploration~~~\mbox{}               & Is there only one dataset?                     & U    \\
&&& direction & \\
top(keywords)    & 3 & (0.7\%)    & Report a subset of the most frequent                       & The most popular license is CC.            & A    \\
&&& concepts & \\
prompt(link)     & 2 & (0.5\%)    & Suggest or request the link to a dataset                           & Send me the link, please.                                & UA \\
\bottomrule
\end{tabular}
\end{table*}

\begin{figure*}
\includegraphics[trim={0.01cm 0 0 0.01cm},clip,width=\linewidth]{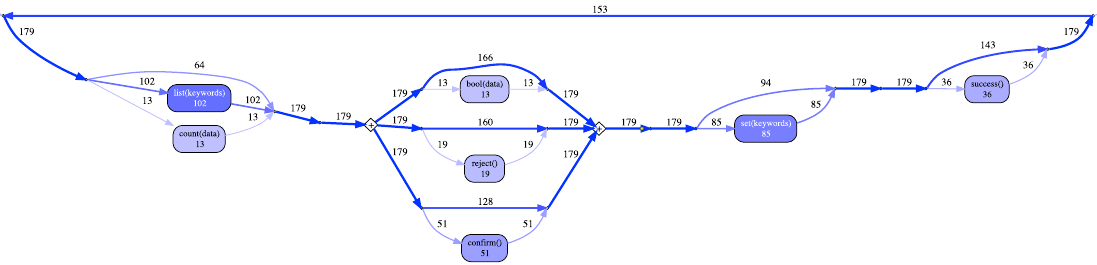}
\centering
\caption{Model of the conversational browsing interaction process extracted from the conversation transcripts. Nodes of the graph correspond to the utterance types described in Table~\ref{dialog_acts} and the arrows between them (transitions) show the direction of the conversation flow. The types may follow each other in a sequence, in parallel (joints marked with ``+'' symbol, meaning that the order varies between the conversations), alternate or form loops. The numbers above the arrows indicate the number of times each transition type occurs in the transcripts. Color intensity indicates relative frequency of the transitions and utterance types in the corpus. (Best viewed in color.)}
\label{fig:conversation_model}
\negskip
\end{figure*}
\end{landscape}

We collect the most frequent patterns of concept types used within the same message and analyze their relations. 
Communicating a subset of entities that belong to the same attribute (or \emph{facet} -- a general category of the attribute) with or without mentioning the name of this attribute as well is the most common pattern we observed. For convenience, we mark such attribute names with boldface: e.g., \textbf{topic}, or \textbf{publisher}. 
Another frequent pattern is listing several attributes or facets within the same message for the Seeker to choose from, e.g., ``I can group the datasets by \textbf{organization}, \textbf{format} or \textbf{topic}.''
A common strategy for the Intermediary is to make an attempt to reduce the subset of items for exploration by prompting the Seeker to select one of the shared attribute values (subset search). We mark values with italics in dialogs: e.g. \textit{education}, or \textit{The City of Vienna}.
When the subset of items is small and more homogeneous, i.e., many datasets have same values of multiple attributes, the Intermediary starts listing values of unique dataset attributes (linear search), such as title, description, or link.

\medskip

In summary, the action space used by the Seeker has three operations:
\begin{itemize}
\item \textit{select} -- provides positive feedback towards one or more of the exploration directions (facets, or attributes), e.g., ``Yes, \textbf{population} sounds interesting'';

\item \textit{skip} -- provides negative feedback towards one or more of the exploration directions, e.g., ``I do not care about the \textbf{data format}'';

\item \textit{prune} -- provides negative feedback towards a subset of items, e.g., ``Is it about \textit{education}? -- No.''
\end{itemize}



The average number of turns per dialogue in our dataset is~5, with the minimum of~1 and the maximum of~14. 
The one turn dialogues consist of answers to direct questions expressing the information need, i.e., the user query.
It usually takes 2--3 turns when the Intermediary also describes the information source before or after answering the user query.
If the Seeker smoothly follows the options offered by the Intermediary, the minimum number of turns for the conversational browsing scenario is~4; it is at least~6 turns if the Seeker rejects some of the options offered by the Intermediary. 
Inefficient strategies leading to an increase in the number of turns required to satisfy the information need include asking general questions and providing an insufficient number of options.

The majority of messages composed by the human Intermediaries contain up to~8 concepts. 
The maximum number of concepts per message was~16, for an extreme case in which the Intermediary listed all available categories. 
The number of concepts per message positively correlates with the performance of the interaction;  Seekers were more likely to find one of the options useful when supplied with more options. 

\section{System Design}
\label{section:design}

We define \textit{conversational browsing} as a collaborative exploration search task with asymmetric roles with uneven distribution of goals and information available to the conversation participants.
One of the conversation participants (the \textit{Intermediary}) has access to the \textit{information model} $I$, while the other participant (the \textit{Seeker}) has access to the \textit{information goal} $G$.
While the goal of the Intermediary is to help the Seeker satisfy the information goal, only the Seeker is in the position to define the goal and/or tell when it is reached.
More formally, the task for the Intermediary is to communicate a subset of $I$ to the Seeker in a sequence of \textit{messages} $M = \langle M_1, \ldots, M_n \rangle$ to form the \textit{knowledge state} $K$ in alignment with $G$ so as to satisfy the success condition $G \subseteq K$.

\begin{figure}[!h]
\centering
\includegraphics[width=.8\textwidth]{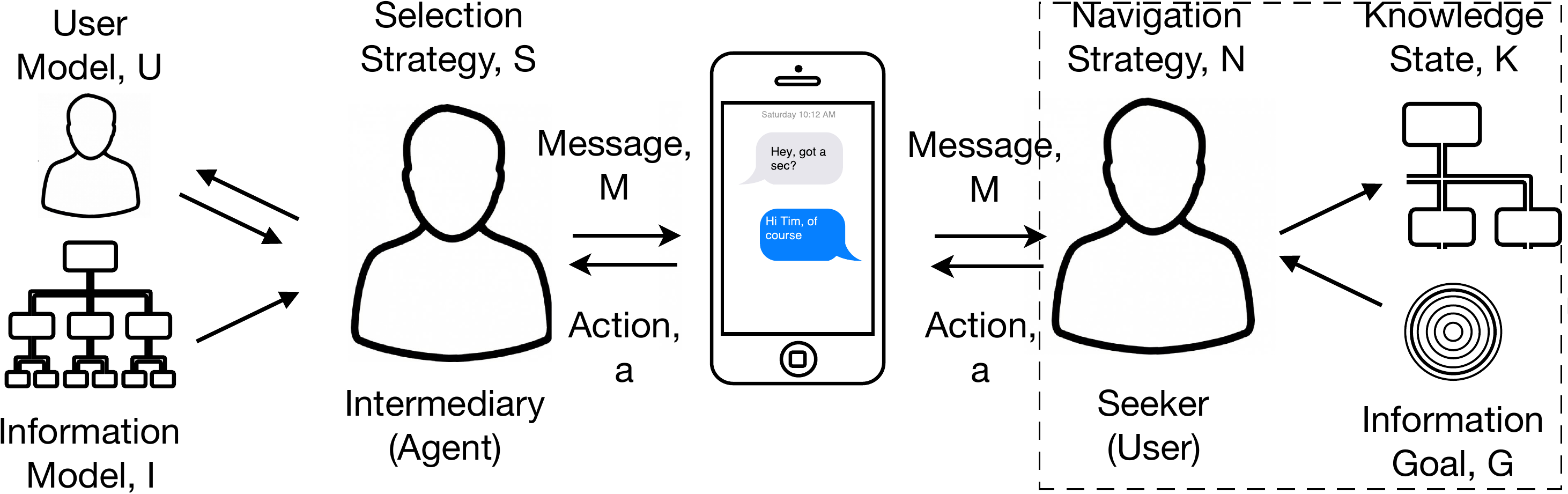}
\caption{Conversational browsing model (CBM). The user model $U$ maintained by the agent is expanded on the right side of the figure consisting of the knowledge state $K$, information goal $G$ and a navigation strategy $N$.}
\label{fig:model}
\end{figure}

We design a \textit{dialogue agent} to play the role of an Intermediary in this task and the \textit{user} to take on the role of the information Seeker. 
A model of the conversational browsing task is illustrated in \figurename~\ref{fig:model}.

\noindent%
The actual knowledge state $K$ and the information goal $G$ of the Seeker are not directly observable by the Intermediary. 
Instead, the Intermediary maintains a \textit{user model} $U$ that reflects the Intermediary's belief about the Seeker's state based on the Seeker's \textit{action} $a$. 
The Seeker chooses one of the actions from the \textit{set of available actions} $a \in A$ using the \textit{navigation strategy} defined as a function $N$, which is a generative process also hidden from the Intermediary.

The Intermediary is assumed to be able to adequately model both the user state $U$ and the information model $I$ in order to choose an optimal \textit{selection strategy} $S$ to compose messages $M = \langle M_1, \ldots, M_n \rangle$. 
The goal in this case is to satisfy the success condition $G \subseteq K$ using a minimal number of messages.

\subsection{User model}
\label{section:user-model}

\textit{Cognitive load} corresponds to the amount of information that working memory can hold and operate on at a unit of time $t$. 
We model cognitive load as a function $L(t)$ that defines the bound on the available cognitive resource of the user, which can represent time, memory, attention span, motivation, patience or user fatigue.

If too much information is presented at once, at time $t$ (that is, if $|M_t| \gg l_t$), the user becomes overwhelmed and much of the information is lost.
Therefore, a na\"{\i}ve brute force selection strategy $S$ that simply pushes the entire information model into a single message is likely to fail according to CLT.
We ground our assumptions about the shape of $L(t)$ in results from cognitive science. 
Various experiments suggest the working memory limit to be close to $7\pm2$~\citep{miller1956magical} objects, or even less~\citep{cowan_2001}. 
While these bounds have been debated, we assume them as a reasonable average $l$ to inform our user model. 

The concept of cognitive load motivates the design of a selection strategy $S$ that takes into account the cognitive resource limitation of the human brain, $L(t)$, to make learning more efficient. 
In particular, it motivates the need for partitioning the message $M$ into a sequence of messages distributed in time $M = \langle M_1, \ldots, M_n \rangle$ with the upper-bound on every message size provided by the cognitive load function such that $|M_{ut}| \leq l_{ut}$.

We assume that the Seeker employs a rational navigation strategy $N$ and is more likely to choose an action $a_{ut}$ that is expected to maximize the knowledge gain with respect to the information goal: $M_{ut+1} \cap G_{ut}$. 
If none of the available actions has any expected value with respect to the information goal $G_{ut}$, the Seeker will choose the action that triggers the default exploration direction selected by the Intermediary based on the structure of the information model.

\subsection{Information model}
\label{section:information-model}

We assume a relational structure of the information source, in which a set of \textit{items} are characterized by a set of \textit{attributes}. 
This grid-like structure is a common data model used in tables and databases across different domains to characterize a group of homogeneous elements, e.g., movies or other products. 

\begin{figure}[t]
\centering
\includegraphics[width=0.7\columnwidth]{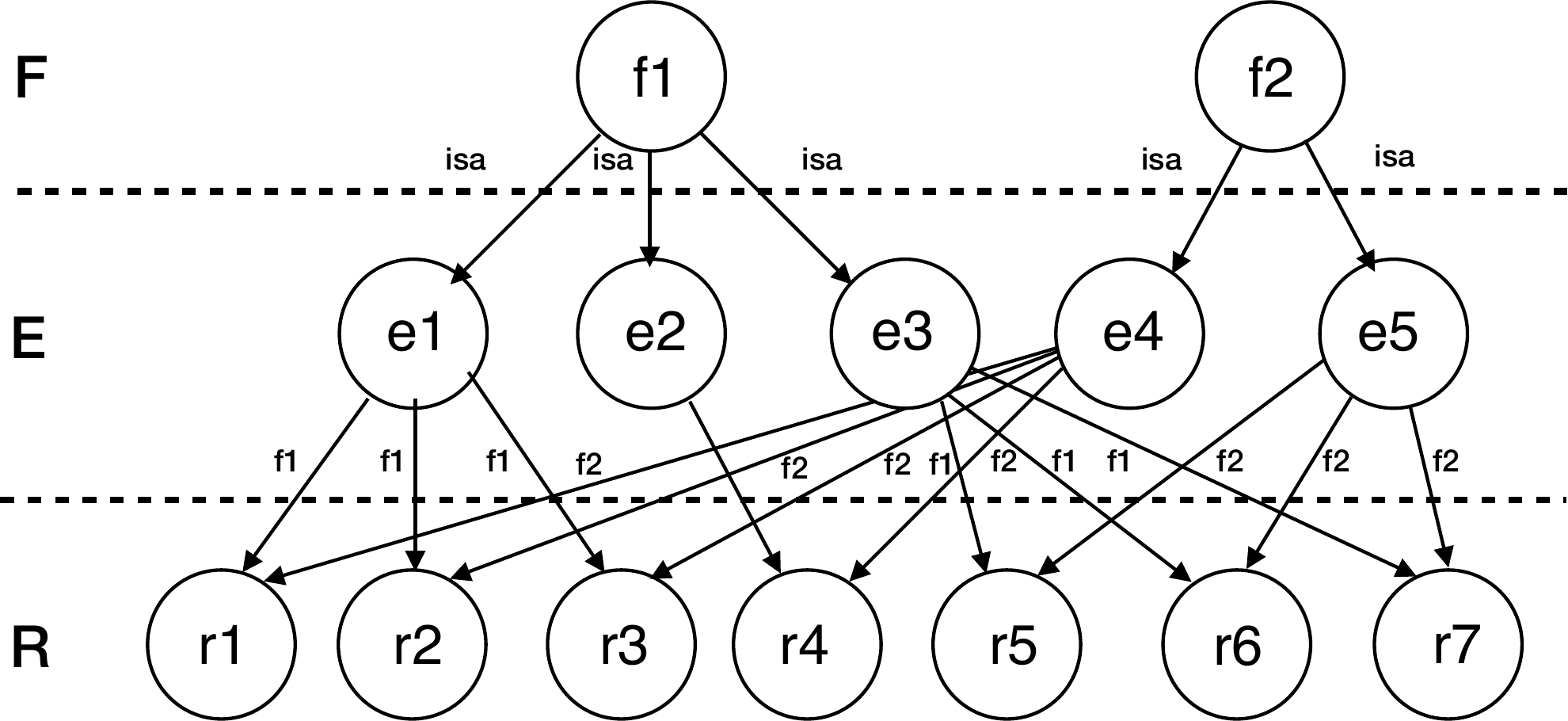}
\caption{Information model with subsets of nodes for attributes $F=\{f1, f2\}$, entities $E=\{e1, \ldots, e5\}$ and items $R=\{r1, \ldots, r7\}$.}
\label{fig:table}
\end{figure}

We represent this data model as a graph in Figure~\ref{fig:table} with three distinct sets of nodes: attributes $F$, entities $E$ and items $R$.
Individual items $R$ correspond to the rows of the table or records in a database; and their attributes $F$ correspond to the columns or facets. 
The intersection of a row $r \in R$ and a column $f \in F$ contains at least one of the \textit{entities} $e \in E$, which corresponds to the value of the attribute $f$ for the item $r$: $f(r) = e$. 
%
Entities can provide \emph{unique} identifiers for specific items, e.g., a name or a URL, or can be \emph{shared} between several items, e.g., location or time dimensions. 
Shared entities provide the structure useful for search and browsing of the collection.

We define a ranking function that calculates the score $v_c$, which allows us to compare the importance of every concept $c$ according to the structure of the information model $I$.

\subsection{Selection strategy}
\label{section:selection-strategy}

The selection strategy $S_{ut}$ takes into account the user model $U_{ut}$ and the information model $I$ described in the previous sections. 
It allows the dialogue agent to construct message $M_{ut}$ to be sent to user $u$ at time $t$. 
The task of the selection strategy $S_{ut}$ is to find the optimal order of the messages. 
The objective is to maximize the amount of information per unit of time respecting the limit of cognitive resource of the user.

The entities are ranked by the number of items they belong to. 
Thus, the unique entities (titles) receive the lowest scores and the most frequent entity in the dataset receives the highest score.
The messages are composed from choosing a subset of top $l$ entities that belong to the same attribute.

\section{System Evaluation}
\label{section:evaluation}

We advance to Step 3.~System evaluation in Fig.~\ref{fig:methodology} and provide
details of the implementation, and two types of evaluation: a user simulation and a user study.
In a user simulation we evaluate the trade-off between the expected number of dialogue turns and the maximum message size and then evaluate the performance of our system in a user study.

All experiments were performed using the dataset downloaded from one of the open data portals, as an information source, that was also used in Step 1.~Data collection.
This dataset contains more than 2,000 items described by 74 different attributes.
Using the conversational transcripts collected in Step 1 we identified a set of 5 attributes that were used by human intermediaries to describe the items: title, license, organization, categorization and tags.

\subsection{User simulation}


\subsubsection{Setup} We evaluate the robustness of our approach to conversational browsing by simulating several user models with different information goals. 
We simulate the Seeker in the following manner.
In every run a new information goal of the Seeker is initialized by picking one of the items (all of its entities) from the database uniformly at random. 
We assume the knowledge state of the Seeker is updated every time the Intermediary sends a message without any loss in the perception channel: $K_{t+1} = K_{t} \cup M_t$. 

The performance metric we used for evaluation was the number of turns the Intermediary needs to satisfy the latent information goal of the Seeker.
A user simulation was applied to tune the cognitive resource capacity $l$, which is the upper bound on the number of concepts the Intermediary can send to the Seeker within a single message $M$: $|M| < l$.
Integer values in the range 3..8 were considered, guided by the related work in cognitive load theory~\citep{miller1956magical,cowan_2001} as well as our own observations drawn from the analysis of the user study in Section~\ref{section:dataset}. 

\begin{table}[ht]
\centering
\caption{Simulation results for different values of the cognitive resource $l$ (500 independent runs).}
\label{simulation}
\begin{tabular}{cccc}
\toprule
\multirow{2}{*}{Cognitive resource $l$} & \multicolumn{3}{c}{Number of turns per dialog} \\
\cmidrule{2-4}
 & Minimum & Average & Maximum \\
\midrule
3 & 5 & 18 & 67 \\
4 & 2 & 15 & 50 \\
5 & 5 & 12 & 80 \\
\rowcolor{Gray}
6 & 2 & 11 & 38 \\
7 & 2 & 9 & 29 \\
8 & 2 & 9 & 27 \\
\bottomrule
\end{tabular}
\negskip
\end{table}

\negskip
\subsubsection{Results}
The results were aggregated across 500 independent simulation runs used for the metric to converge and are listed in Table~\ref{simulation}.
The minimum number of turns to satisfy the simulated information need is~2, when the first message contains an entity that  is able to uniquely identify the item $G$ and the second message contains all the entities that belong to the item $G$. 
The average and maximum numbers of turns required to reach $G$ monotonically decrease with the increase of the parameter $l$.
The results show that a greedy heuristic maximizing the out-degree in our information model performs reasonably well in the selection strategy for conversational browsing but is sensitive to the value of parameter $l$. 
\noindent%
Based on these estimates we chose the value of hyperparameter~$l = 6$ as an estimate of the cognitive resource limit in our user model that determines the maximum number of concepts per message. 
In this case, a simulated user will require~11 actions, on average, to reach any item in the dataset using our conversational browsing system.

\subsection{User study}

\begin{figure}[t]
\includegraphics[clip,trim=0mm 0mm 0mm 0mm, width=\columnwidth]{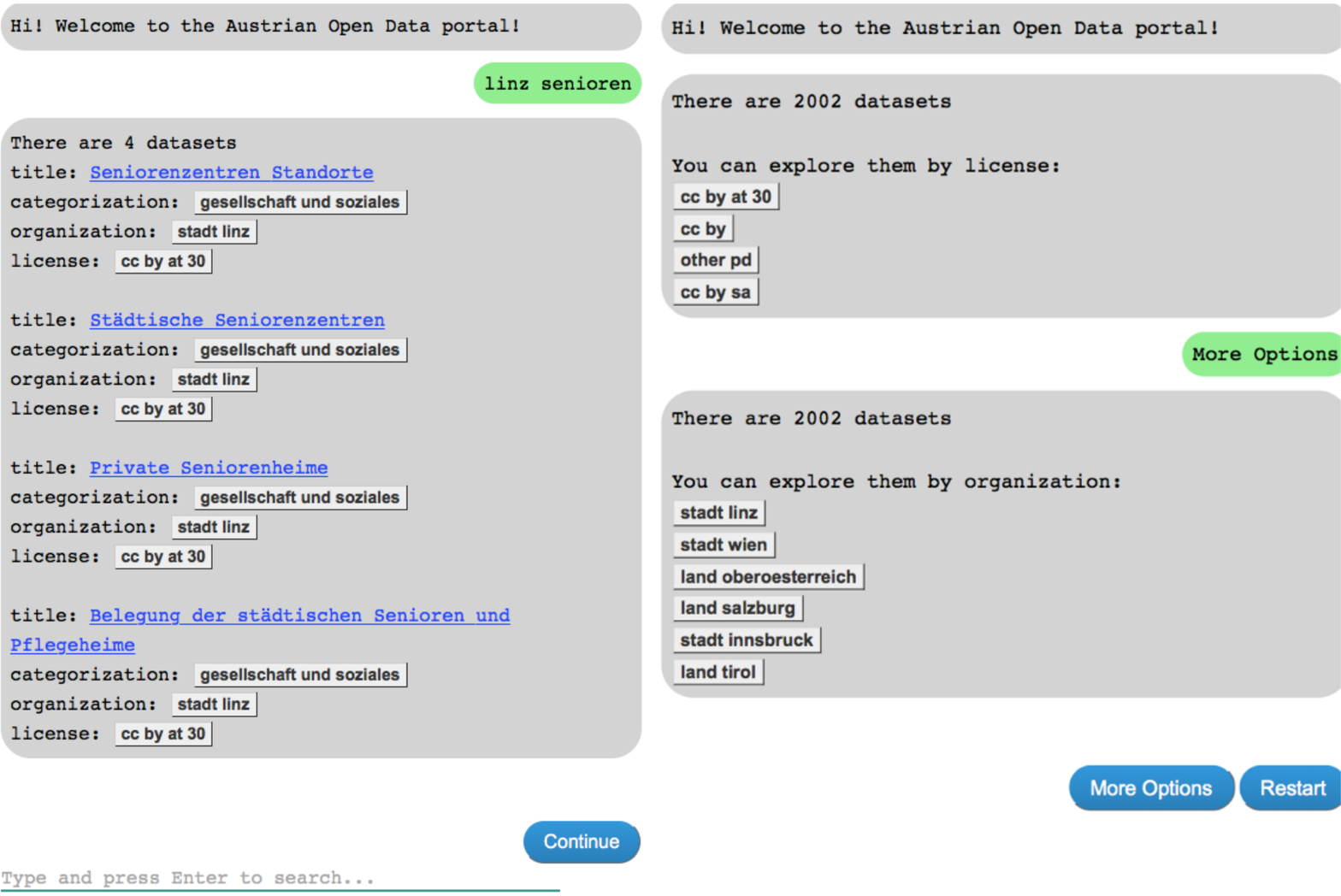}
\caption{Two types of interaction with the system: conversational search (left) versus conversational browsing (right).}
\label{fig:screenshots}
\negskip
\end{figure}

For the user study we designed two conversational interfaces (see \figurename~\ref{fig:screenshots}).
The first one (left) provides a typical conversational search functionality: the user query is used to produce a ranked list of the matching items retrieved from the index.
The alternative interface (right) implements conversational browsing functionality by interactively revealing the subsets of the most discriminative attributes based on the user feedback.

The goals of browsing, as an information-seeking strategy, can vary from general collection understanding and learning to exploratory search~\citep{bates1989design}.
It is very challenging to evaluate the success of learning and the level of understanding.
We focus on the later goal of exploratory search instead, defined as an ability to discover relevant information via browsing.
Moreover, in this setup we can directly compare the results achieved using the proposed conversational browsing interface on the same search tasks that can be completed using the query-based conversational search interface, which serves us as a baseline.
\subsubsection{Setup} The total of 24 participants took part in the experimental evaluation of our conversational browsing approach.
The volunteers were recruited among the university students and none of them participated in our data collection study,
All participants had previous experience with basic web search interfaces, such as keyword- and faceted search, but no previous experience with the repositories, web-sites or our conversational interfaces used in this user study.


Each participant filled out a questionnaire that included a competency question for assessing the prior domain knowledge, accommodated search results for two search scenarios, and asked for participant's feedback in the end of the experiment. 
In this way we collected two types of feedback: 
\begin{enumerate*}[label=(\arabic*)]
\item subjective feedback by the participants reflecting on their experience using the systems; and \item objective feedback reflecting the average performance on the search task using different systems\end{enumerate*}.
%

%
\begin{table}[ht]
\centering
\caption{User study success rates: proportion of the users who successfully completed the search and browsing tasks.}
\label{user_study}
\begin{tabular}{ l c c c c }
\toprule
 & \multicolumn{2}{c}{Task}                             &       \\ 
\cmidrule{2-3}
System       & (1) Immigration Vienna~~~ & (2) Retirement Linz &~~~& Total \\ 
\midrule
Search & 0.33                  & 0.08               && 0.21  \\ 
Browse~~~\mbox{} & 1.00                  & 0.33               && 0.67  \\
Total  & 0.67                  & 0.21               &       \\ 
\bottomrule
\end{tabular}
\negskip
\end{table}

To design the sample information seeking scenarios we picked two items from the dataset at random and formulated the tasks for the user based on these items. 
We carefully phrased the search task so as to reflect the vocabulary mismatch problem, which often occurs in real-world settings, by rephrasing some of the keywords in the title and other attributes of the target items:

\begin{enumerate}
\item \emph{Population by country of birth since 2011 municipal districts Vienna:\footnote{\url{https://www.data.gv.at/katalog/dataset/0a0f2617-3609-42ca-97bc-2f8a8be98cbf}} locate datasets that can provide information about immigration in Vienna}; and
\item \emph{Private retirement homes:\footnote{\url{https://www.data.gv.at/katalog/dataset/8421a66f-dc80-4bd3-8253-de532bc5b67c}} locate datasets that can provide important information especially for the older generation of adults living in and around Linz}.
\end{enumerate}


We evaluate performance on the tasks using a success rate that corresponds to the number of participants who manage to successfully complete the task by finding at least one of the correct datasets and analyzing the number of turns it took users to complete the task to compare it with the expected performance from our simulation. 
It took our user simulation between~5 and 9~turns to reach the item for every item in the pool of correct results, with~6 and 8~turns on average for the different tasks.

\negskip
\subsubsection{Results} 
\begin{figure}[t]
\includegraphics[width=\columnwidth]{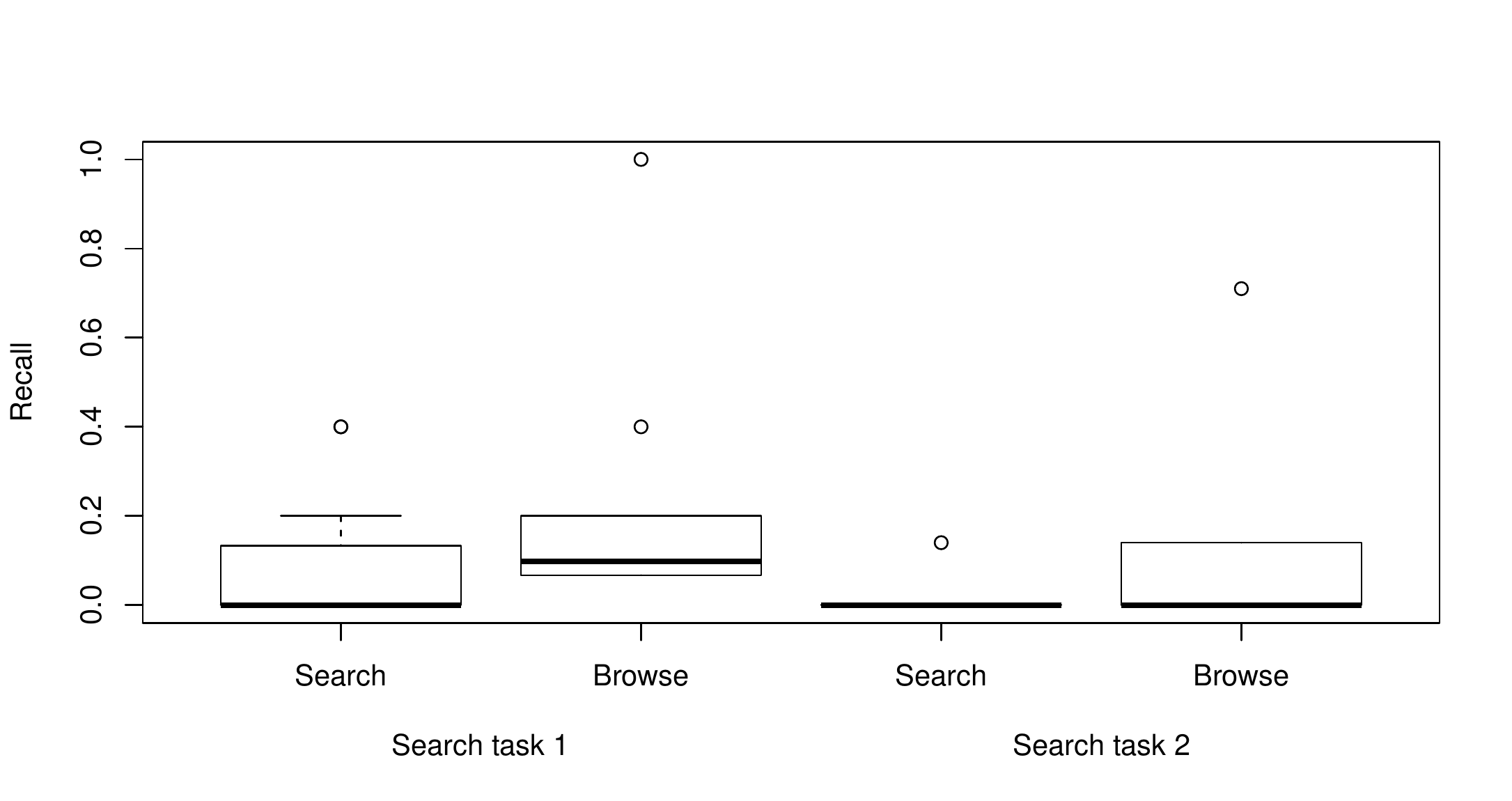}
\caption{Distribution of recall results from the user study.}
\label{fig:ncorrect}
\end{figure}

On average, participants performed better using our conversational browsing system in comparison with the basic search functionality. 
Only 5~out of 24~participants managed to complete the search task using the baseline system, in comparison with a success rate of more than 50\% for the conversational browsing system (16~out of 24). 
Table~\ref{user_study} displays the task success rate, i.e. the ratio of users who successfully completed the task.
The recall for both tasks was also higher for the conversational browsing system than for the search interface (see \figurename~\ref{fig:ncorrect}).

The second task (``Retirement Linz'') turned out to be much harder to complete than the first one (``Immigration Vienna'') as evident from the difference in success rates. 
All participants succeeded when using conversational browsing for the first task, and a third of them -- for the second task, in comparison with the third of the participants for the first task and a single person only for the second task, when using the baseline search system. 

We created a pool of results submitted by participants for each of the tasks to enrich our subset of results marked as correct.
Two independent annotators marked correct answers in the pool with an inter-annotator agreement of 0.95 and resolved the disagreements by discussing the content of the datasets.
Statistics for each of the tasks completed via the browsing interface is summarized in Table~\ref{task_stats}, where the number of all unique results as submitted by the users corresponds to the pool, and the number of correct results are the ones marked as relevant with respect to the task by the annotators.
For more difficult task the fraction of incorrect results submitted is higher.
Also the users took more dialogue turns and restarts to complete the more difficult task (16 versus only 2 for the simpler task).
The number of turns predicted in the simulation is also higher for the more difficult task, but the gap is much bigger for the real users, which is likely due to the restarts the users take when they are not sure that they are navigating in the right direction.

\newcommand{\SingleD}[1]{\phantom{0}#1}

\begin{table}[t]
\centering
\caption{Statistics for the tasks: size of the result pools and average number of turns.}
\label{task_stats}
\begin{tabular}{ l@{}c  c }
\toprule
\multirow{2}{*}{Statistics}                    & \multicolumn{2}{c}{Task}                              \\ 
\cmidrule{2-3}
                              & (1) Immigration Vienna~~~ & (2) Retirement Linz \\ 
\midrule                              
All results                   & 19                    & 28                 \\ 
Correct results               & 15                    & \SingleD{7}                  \\ 
\#Turns simulated & \SingleD{6}                     & \SingleD{8}                  \\ 
\#Turns user study~~~\mbox{} & \SingleD{8}                     & 21                 \\ 
\#Restarts                    & \SingleD{2}                     & 16                 \\ 
\bottomrule
\end{tabular}
\negskip\negskip
\end{table}

\noindent%


\begin{figure}[!t]
\includegraphics[width=\columnwidth]{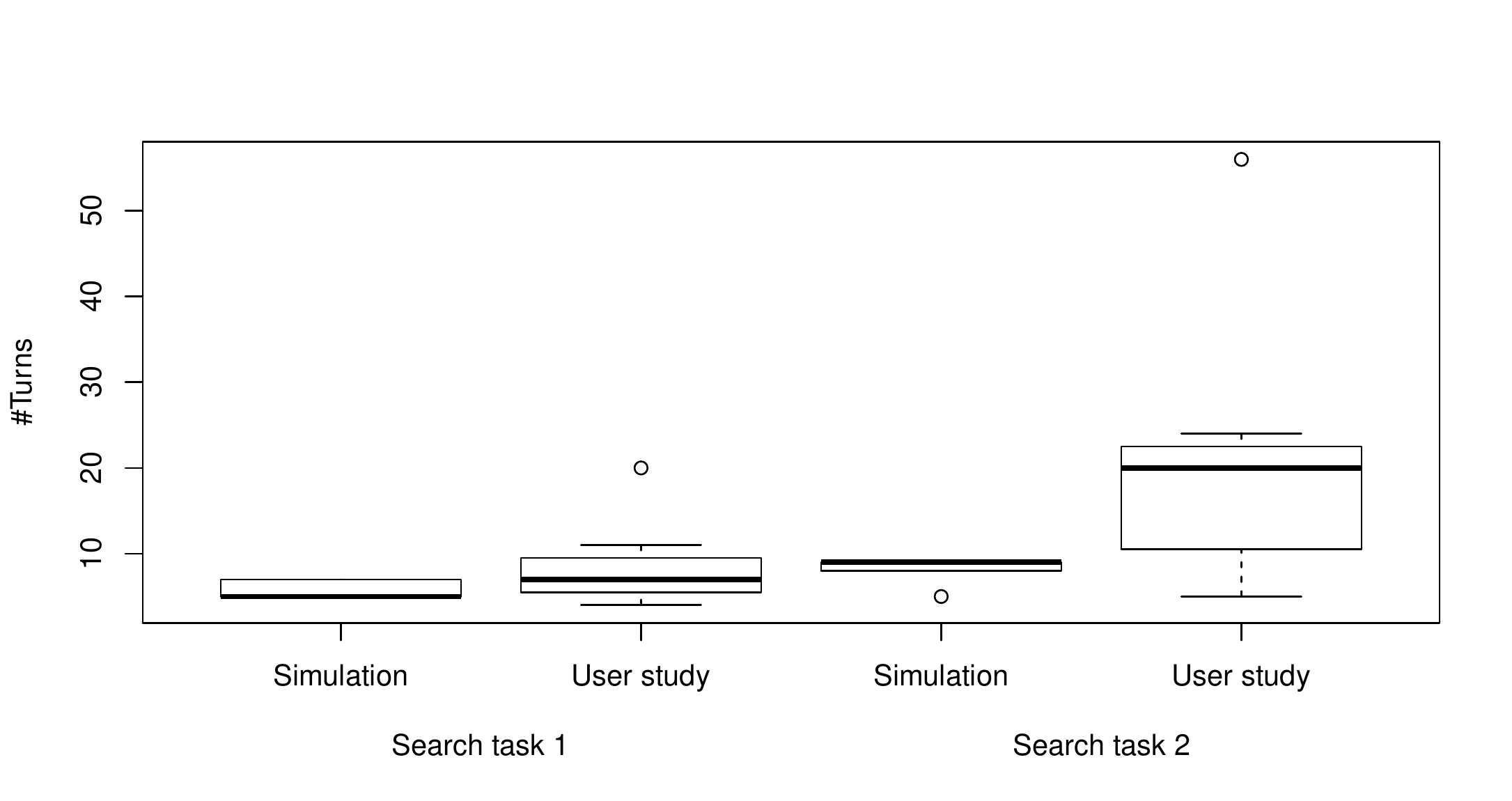}
\caption{Number of turns taken to complete the conversational browsing task (simulation and human participants).}
\label{fig:nturns}
\negskip\negskip
\end{figure}

\figurename~\ref{fig:nturns} shows that the items ranked high with our ranking function, i.e. the items with the most frequent attributes, are much easier to find than the items with less frequent attributes.
The average number of turns for the first search task was~8 (vs.\ 6~turns in the user simulation) and 21~for the second task (vs.\ only 8~in simulation).

\subsection{Discussion}

The user study results showed that the conversational browsing functionality helps users to mitigate the vocabulary mismatch problem and find relevant information, even in the case of limited domain knowledge.
We observed a striking difference in the search performance: the success rates of the browsing interface are three times higher than the query-based search interface on the same search tasks (see Table~\ref{user_study} and \figurename~\ref{fig:ncorrect}).
Among the positive feedback for the conversational browsing system were the ease of use and clarity, and the possibility to explore the data when the search criteria are not clear.

The majority vote, however, showed the opposite result in favor of the baseline: 70\% of the study participants preferred the conversational search interface rather than conversational browsing, when they were explicitly asked for their subjective feedback with the question ``Which system did you like more?''.
We attribute this result to several factors:
(1) all participants had previous \textbf{experience} with query-based search engines, which are similar to the functionality and the interaction type, which our conversational search interface provides;
(2) the participants received a description of the information \textbf{goal}, which they could use to formulate the search query faster than browse the entire collection; 
(3) finally, the participants were not able to adequately assess their search performance since the correct results were not provided, which in turn lead to the \textbf{misconception} about the system performance that likely influenced the preference choice.

More experiments are needed to evaluate usability and integration of browsing components into the conversational search interfaces.
Also, the selection strategy should be able to integrate alternative ranking functions beyond the information-theoretic objective only.
The results of the user study showed that some of the attributes identified as highly discriminative for the given dataset, such as data license, could not help users to decide on relevance.
User preferences, such as perceived attribute relevance, can be either collected in a separate survey or a user feedback form, or harvested from the logs of the conversational search system directly.


Our experimental results support previous findings and claims that dialogue systems can be an effective instrument for information retrieval also without the need to explicitly formulate the query, which can be especially relevant in the situations promoting serendipitous discovery and general collection understanding~\citep{oddy1977information}.
We complement previous work in this direction by providing an extensive description of the approach we used and various aspects of its evaluation extended with the analysis of the challenges that arise in the design and evaluation of this kind of systems.

\section{Related Work}
\label{section:related}

% ...

\subsection{Interactive retrieval}
Interactive information retrieval interfaces are designed to provide continuous support on different stages of the information seeking process~\citep{belkin2001iterative}.
Common interactive IR techniques developed to assist users in query formulation include term and query suggestion~\citep{Harman:1988:TIQ:62437.62469,koenemann1996case,kelly2009comparison}.
The task is then formulated as an optimization problem to rank and select a subset of options, e.g., terms, to show to the user, which is motivated by limits in cognitive load and window size~\citep{Harman:1988:TIQ:62437.62469,ruthven2003re,kumaran2008effective}

The Text REtrieval Conferences (TREC) Interactive Track was an initiative that followed the iterative approach to the design of an interactive IR interface with evolutionary design/evaluation cycles over the course of four consecutive years~\citep{belkin2001iterative}. 
The major findings, which led to several design improvements, were reported to be the following:
(1)~users prefer transparency and control over the search mechanisms;
(2)~one of the crucial points in adopting the functionality is ease-of-use, i.e., the search interface has to be simple and intuitive; and
(3)~collection-based terms are perceived as more useful than the terms derived from relevance feedback.
In this paper we build upon these results and propose the design of a new interface that satisfies all the requirements listed above.
It is reduced to a single chat window and the set of available options is unveiled to the user interactively based on the choices made in the previous steps.
This interface provides the functionality for interactive retrieval via collection browsing without the requirement to formulate a query.
Similar to previous work~\citep{wambua2018interactive}, we exploit the properties of the document collection to guide the user along various facets ranked by the information gain criteria and further extend it in the context of conversational browsing interfaces.


\subsection{Dialogue systems}

Due to recent advances in speech recognition and natural language generation technologies conversational interfaces experience a surge of interest from both industry and academia~\citep[see, e.g.,][]{scai,DBLP:conf/acl/DhingraLLGCAD17,DBLP:journals/dad/WilliamsRH16a}.
There are two major types of dialogue system application considered now: chit-chat dialogues and task-oriented dialogues~\citep{DBLP:journals/ftir/GaoGL19}.
The idea of an automated system being able to communicate by means of a natural language goes back to the beginning of computing.
The seminal paper by \citet{alan1950turing}, which introduced a human-machine dialogue as a test for artificial intelligence, is still regarded as an ambitious goal and drives the development of systems capable to hold an open-domain (chit chat) conversation.
In practice, this kind of conversational systems are being used predominantly for entertainment~\citep{DBLP:journals/corr/abs-1801-03604}, in more radical cases even aiming to replace a human companion\footnote{\url{https://replika.ai}}~\citep{DBLP:journals/corr/abs-1712-05626}.

Another direction for conversational systems are task-oriented dialogues, in which one of the main measures of conversation success is the task completion ratio that makes evaluation more straight-forward.
Classic examples of task-oriented dialogue systems are in the restaurant reservation and trip planning domains~\citep{DBLP:journals/dad/WilliamsRH16a}.
In this case the dialogues that the system is able to support are more specialized and domain-specific, in comparison with their chitchatting counterparts.
Task-oriented dialogues rely on a domain specification, which can be defined in terms of an ontology, a table or a set of annotated dialogue samples that provide a frame for linking slots and intents to possible replies.
This ontology enumerates all concepts and attributes (slots) that a user can specify or request information for~\citep{DBLP:conf/acl/MrksicSWTY17}. 
The dialogue models are then designed to jointly perform the tasks of parsing the input utterances, slot matching/filling and belief state tracking~\citep{williams2013dialog}.


Conversational browsing are conceptually different from a task-oriented dialogue, where an agent tries to pin-point an item or an information subspace relevant to the user's query~\citep{DBLP:conf/chiir/RadlinskiC17}.
In this respect, conversational browsing is hard to optimize, since there is no single correct answer.

Conversational browsing is similar to the information presentation subtask of a dialogue system designed to optimise display of available options to a user~\citep{DBLP:conf/eacl/DembergM06,DBLP:journals/taslp/RieserLK14}.
However, conversational browsing does not assume an initial user query, i.e. available options always equate to the whole information space.
With the amount of information that can be potentially communicated to a user getting larger a major design challenge arises with respect to taking in account cognitive limitations of the human brain for partitioning the information space into messages and using structural properties of the information space to allow a more efficient traversal.


\subsection{Conversational search}
Lately, conversational agents and conversational search systems are becoming increasingly popular~\citep{DBLP:conf/chi/VtyurinaSAC17}.
So far, however, such systems mainly focus on question answering and simple search tasks, those that are to a large extent solved by web search engines.
Since the early 1960s there have been many efforts in the database community to support natural language queries by translating them into structured queries~\citep[see, e.g.,][]{green-automatic-1963,woods-lunar-1977,bronnenberg-question-1980}.
Most of the on-going work in conversational search is still focused on the question answering task, which is an important interaction type in the context of an information seeking conversation but not the only one the conversational system has to provide support for~\citep{DBLP:journals/corr/abs-1812-10720}.
We argue that conversational agents and search systems should also support exploratory search.


One work that had a profound impact on the development of information seeking models and, in particular, on the concept of the anomalous state of knowledge (ASK)~\citep{belkin1982ask}, was the paper describing a dialogue system called THOMAS~\citep{oddy1977information}.
The main idea manifested in THOMAS, and later expanded and more thoroughly formalized in ASK, was that often the user is ``not able to formulate a precise query, and yet will recognize what he has been looking for when he sees it''~\citep{oddy1977information}.
The paper does not detail the implementation of the proposed approach but reports that it achieves on-par performance in comparison with a standard query-based search system.

In this paper we reincarnate the idea underlying THOMAS to validate the hypothesis that a browsing-based conversational system is able to satisfy the information needs of a user without the burden of query formulation.
We also provide the evaluation results of this concept in a user study.

\section{Conclusion}
\label{section:conclusion}

We introduced a novel conversational dataset illustrating the asymmetric collaborative information-seeking scenario, in which an Intermediary, having access to an information source, plays a pro-active role by interactively revealing and dynamically adjusting the possible exploration directions based on the feedback from an information Seeker.
This scenario, which we cast as the \emph{conversational browsing task}, is appropriate when the Seeker is not sufficiently familiar with the domain of interest to be able to formulate their information need as a concise search query, or prefers to explore the available options.

We proposed a formalization of conversational browsing as an interactive process in which the Intermediary guides the Seeker in discovering the relevant attributes (facets) and filtering conditions (entities) to single out a subset within the information source that contains the information goal of the Seeker. 
Our experiments indicate that conversational browsing is a viable paradigm able to mitigate the challenges in query formulation and assist users in conversational search.


The dataset and the model that we proposed indicate much broader implications for conversational system design than we could utilize in the first set of experiments.
Despite these limitations, we believe that our initial results showcase conversation browsing as a useful component for conversational search, which is able to complement the already established question answering task and encourage development for the set of more advanced interaction patterns with a dialogue system.

While similar ideas were already discussed decades ago~\citep{oddy1977information}, they were abandoned at the time, not matched by adequate technology for natural language understanding~\citep{belkin2016people}.
We believe that it is time to revisit these ideas.
The combination of novel techniques for semantic parsing and information retrieval with more advanced information-seeking models constitutes a promising direction for future work.

More recently we experimented with designing a single conversational interface, which integrates both browsing and search functionality that naturally complement each other~\cite{DBLP:conf/esws/KeynerSV19}.
It allows the user to reduce the search space using a custom query and browse the subset of search results; or browse to get an overview of the content and domain vocabulary and afterwards formulate a query based on the acquired knowledge of the collection.
Enabling seamless transitions between these two interaction modes is an important direction for future work.

The conversational browsing functionality should be also further extended beyond exploring the metadata model only.
We would like to enable exploratory search over the dataset content that can complement question answering approaches over tables and knowledge graphs~\cite{DBLP:conf/i-semantics/VakulenkoS17,DBLP:conf/cikm/VakulenkoGPRC19}. 
Therefore, an important direction for future work is to extend the information model to handle arbitrary graph structures beyond a single metadata table.
Last but not least, there is room for a learning component able to learn from the interactions with users and incorporate the user feedback to improve the overall performance as well as maintain personalized user models.

\bibliographystyle{plainnat}
\bibliography{references}

\end{document}